\documentclass[aps,prl,twocolumn,showpacs]{revtex4-1}
\usepackage{amsmath}
\usepackage{graphicx,epsfig}
\usepackage{bm}

\renewcommand{\v}[1]{{\bf #1}}
\newcommand{\be}{\begin{equation}}
\newcommand{\ee}{\end{equation}}
\newcommand{\bd}{\begin{displaymath}}
\newcommand{\ed}{\end{displaymath}}
\newcommand{\ba}{\begin{eqnarray}}
\newcommand{\ea}{\end{eqnarray}}
\newcommand{\nn}{\nonumber \\}
\newcommand{\bpm}{\begin{pmatrix}}
\newcommand{\epm}{\end{pmatrix}}

\begin{document}

\title{Emergence of Chiral Magnetism in Spinor Bose-Einstein Condensate \\ with Rashba Coupling}

\author{Xiao-Qiang Xu}
\affiliation{Department of Physics, BK21 Physics Research Division,
Sungkyunkwan University, Suwon 440-746, Korea}
\author{Jung Hoon Han}
\email[]{hanjh@skku.edu}
\affiliation{Department of Physics, BK21 Physics Research Division,
Sungkyunkwan University, Suwon 440-746, Korea}


\begin{abstract} Hydrodynamic theory of the spinor BEC condensate
with Rashba spin-orbit coupling is presented. A close mathematical
analogy of the Rashba-BEC model to the recently developed theory of
chiral magnetism is found. Hydrodynamic equations for mass density,
superfluid velocity, and the local magnetization are derived. The
mass current is shown to contain an extra term proportional to the
magnetization direction, as a result of the Rashba coupling.
Elementary excitations around the two known ground states of the
Rashba-BEC Hamiltonian, the plane-wave and the stripe states, are
worked out in the hydrodynamic framework, highlighting the
cross-coupling of spin and superflow velocity excitations due to the
Rashba term.
\end{abstract}
\maketitle

A recent excitement in the cold atom physics is the newly found
ability to engineer spin-orbit-type interactions among the spinor
$^{87}$Rb atoms~\cite{spielman-nature}. The effort is part of a
broader theme to create synthetic gauge field environment for cold
atoms~\cite{dalibard-review} and currently forms one of the most
exciting branches of cold atom research. Several theoretical papers
appeared dealing with the possible phases of spinor Bose-Einstein
condensate (BEC) in the presence of spin-orbit interaction of the
type $\sim \Psi^\dag \v F \cdot \v p \Psi$, for an appropriate spin
operator $\v F$, the spinor $\Psi$, and momentum $\v p = -i\hbar\bm
\nabla$~\cite{zhai,rashbaBEC}. Due to the similarity
of this term to the Rashba coupling $\sim \sigma_x p_y - \sigma_y
p_x$ (which can be mapped to $\sigma_x p_x + \sigma_y p_y$ with
suitable spin rotation) in solid state physics~\cite{rashba}, spinor
BECs containing such interaction will be called Rashba-BECs. The
Rashba coupling contains one space gradient while the usual
kinetic energy has two. The competition between the two energies results
in a length scale, as manifested for instance in the striped
superfluid state in certain interaction parameter regimes of the
Rashba-BEC Hamiltonian~\cite{zhai}. The influence of confining trap
on the Rashba-BEC was also considered~\cite{trap}. In another vein,
the rotation effect in a harmonically trapped Rashba-BEC was
studied in several papers~\cite{rotatingRashba}.
Fast-rotating limit of the Rashba-BEC is expected to yield
non-Abelian Landau level structures with the exotic possibility of a
many-body state characterized by fractional
statistics~\cite{burrello}.

Most theoretical works so far on Rashba-BEC have been numerical
studies of the ground states. Recently, an interesting
possibility of fractionalized vortex excitation in the striped
superfluid state and associated thermodynamics of Kosterlitz-Thouless
phase transition were considered in Ref.~\cite{cao}. Still, the
general characterization of the Rashba-BEC system in hydrodynamic
variables and identification of proper conservation laws are lacking.
In Ref.~\cite{biao-wu} the authors studied
fluctuations of the so-called plane-wave state within the
framework of
the time-dependent Gross-Pitaevskii (GP) equation directly, while the  discussion
of more general states in spin-orbit coupled BECs could be found in Ref.~\cite{Barnett}.
In this
paper, in the tradition of formulating superfluid phenomena in
hydrodynamic variables~\cite{khalat}, we derive conservation laws of
mass density ($\rho$) and mass current ($\v J$) for Rashba-BEC
system. In addition, Euler equation for the superflow velocity $(\v
v)$ and Landau-Lifshitz equation for the spin vector ($\v n$) dynamics are
obtained. These are generalizations of the hydrodynamic formulation
carried out for non-Rashba spinor BEC case in
Refs.~\cite{lamacraft,refael}.
As will be shown, spin-orbit coupling introduces several new features
missed in previous works such as the rigorous mathematical mapping to
chiral magnets and the modification in the meaning of the current density.
 The set of equations are then applied
to study coupled fluctuation dynamics of the hydrodynamic coordinates
for plane-wave and striped superfluid states of Rashba-BEC
Hamiltonian.
Hydrodynamic theory provides a useful alternative to the
Bogoliubov approach to study
fluctuations~\cite{demler} and our calculations illustrate the coupled nature
of spin and superflow velocity in the excitation spectrum.


We begin by writing down the single particle Hamiltonian of
Rashba-BEC system as
\ba H_0 &=& {1\over 2} \int [ (\v p +\kappa \v F ) \Psi]^\dag \cdot
[ (\v p +\kappa  \v F ) \Psi] , \label{eq:BEC-Rashba} \ea
where $\kappa$ describes the strength of Rashba spin-orbit coupling.
We will assume $\hbar=m=1$ throughout the paper. The task at hand is
to transform the Rashba-BEC Hamiltonian $H_0$ using hydrodynamic variables.
For spin-1/2 case in particular, we have $\v F = \bm
\sigma$ ($2\times 2$ Pauli matrices) and one can decompose the two-component spinor field $\Psi$
as~\cite{lamacraft,refael}
\ba \Psi = \bpm \psi_\uparrow \\
\psi_\downarrow \epm = \psi \v z , ~~
\v z =  \bpm \cos(\theta/2) e^{-i \phi /2 } \\
\sin (\theta/2) e^{i\phi/2} \epm , \label{eq:CP1}
\ea
where $\psi = \sqrt{\rho} e^{i\xi}$ with $\xi$ being the overall
phase of the spinor condensate. The CP$^1$ field $\v z$ appears
in the theory of nonlinear $\sigma$-model (NL$\sigma$M). The
relevant hydrodynamic variables are defined in terms of the spinor
field as $\rho = \Psi^\dag \Psi$ and $\v n = \v z^\dag \bm \sigma \v z
= \Psi^\dag \bm \sigma \Psi/\rho = (\sin\theta \cos\phi,\sin\theta \sin\phi, \cos\theta)$.
 Superflow velocity is defined as
$\v v= \v a_c - \v a_s$, where the two vector potentials are $\v a_c
= -i e^{-i\xi} \bm \nabla e^{i\xi} = \bm \nabla \xi$ and $\v a_s = i
\v z^\dag \bm \nabla \v z$. Certain
identities can be derived relating terms in the GP Hamiltonian to the
hydrodynamic expressions as~\cite{lee-kane,han}:
\ba \label{eq:spin-half-identity}
( \v p \Psi)^\dag \cdot ( \v p
\Psi ) &=& \rho \Bigl({1\over 4} \sum_i
\partial_i \v n \cdot
\partial_i \v n + \v v^2 \Bigr) + (\bm \nabla \sqrt{\rho})^2, \nn
\Psi^\dag \bm \sigma \cdot ( \v p \Psi ) + h.c.  &=& \rho\Bigl(  \v
n \cdot (\bm \nabla \times \v n) + 2 \v n \cdot \v v \Bigr) .
\ea
The Rashba-BEC Hamiltonian (\ref{eq:BEC-Rashba}) is accordingly
transformed to $H_0 = H_1 + H_2 + H_3$ where
\ba
&& H_{1} = {\rho \over 2} \Bigl[ {1\over 4} \sum_i
\partial_i \v n \cdot \partial_i \v n \!+\! \kappa \v n \cdot (\bm
\nabla \times \v n)  \Bigr], \nn
&& H_2 = {1\over 2} \left(\bm \nabla \sqrt{\rho} \right)^2 , \nn
&&  H_3 =  {\rho \over 2} ( \v v \!+\! \kappa \v n )^2  .
\label{eq:H-transformed}
\ea
A similar exercise for spin-1
operator $\v F$ and ferromagnetic spinor wave function~\cite{ho98}
yields the same hydrodynamic Hamiltonian as
Eq.~(\ref{eq:H-transformed}), only with 1/4 in $H_1$ being replaced by
$1/2$. The analysis carried out below therefore applies equally well
to both two- and three-component spinor BECs with Rashba coupling.

The key new element in the spin dynamics ($H_1$) introduced by
nonzero $\kappa$ is $\v n \cdot (\bm \nabla \times \v n)$. In solid
state magnetism, such a term is known as the Dzyaloshinskii-Moriya
(DM) interaction. Taken together with the NL$\sigma$M term, the
ground state of the spin Hamiltonian $(1/4)\sum_i
\partial_i \v n \cdot \partial_i \v n + \kappa \v n \cdot (\bm \nabla \times \v n)$
is the well-known spiral spin phase with the modulation wave
vector $|\v q | = 2 \kappa$, and the spins rotating in the plane
perpendicular to $\v q$. Here we have $1/4$ for spin stiffness
$J$ in magnetism, and $\kappa$ as the ratio of DM energy $D$ over
$J$. As we will show shortly, the stripe state~\cite{zhai} carries
precisely such spin structure. In the incompressible limit,
assuming $\v v + \kappa \v n =0$ to minimize $H_3$, we find the perfect
match of the Rashba-BEC Hamiltonian to that used in the study of spiral magnetism~\cite{han}.
In the NL$\sigma$M+DM spin model, a textured spin phase called the Skyrmion crystal was
identified in the presence of weak magnetic field or spin anisotropy (See Ref.~\cite{han} and articles cited therein). We
believe it is no coincidence that a very similar Skyrmion lattice
was found in two recent theoretical studies of the phase diagrams of Rashba-BEC
Hamiltonian in a trap~\cite{trap}.

Next we present the dynamical equations of motion for $\rho$, $\v
v$, and $\v n$, based on the Hamiltonian $H=H_0 + H_\mathrm{I}$
where $H_0$ is given in Eq. (\ref{eq:BEC-Rashba}) and
$H_\mathrm{I}$, the interaction, is given by
\ba H_\mathrm{I} = {1\over 2} \int  [g_c (\rho_\uparrow +
\rho_\downarrow )^2 - g_s (\rho_\uparrow - \rho_\downarrow )^2 ] \ea
for spin-1/2 case. Here $\rho_\sigma$ ($\sigma = \uparrow, \downarrow$) refers to the density of the
spin-$\sigma$ component. Note that
$g_c=(g_{\uparrow\uparrow}+g_{\uparrow\downarrow})N/2$,
$g_s=(g_{\uparrow\downarrow}-g_{\uparrow\uparrow})N/2$, with
$g_{\uparrow\uparrow}$ and $g_{\uparrow\downarrow}$ being the intra-
and inter-component $s$-wave interactions, respectively, and $N$ is the total
particle number. It is assumed that
$g_{\uparrow\uparrow}=g_{\downarrow\downarrow}$ and $g_{\uparrow\downarrow}=g_{\downarrow\uparrow}$. We state the
equations first, and discuss briefly methods of derivation later.
\\

\textit{(1) Mass conservation equation}: \ba \partial_t \rho + \bm
\nabla \cdot [\rho (\v v + \kappa \v n ) ] = 0.
\label{eq:massConservation} \ea
This is the generalization of the usual mass conservation equation to
the case with
nonzero Rashba coupling parameter $\kappa$. The combined expression
$\rho (\v v + \kappa \v n) = \v J$ serves as the mass current
density. The relationship can be expected from taking a functional
derivative of the hydrodynamic Hamiltonian $H_0$ given in Eq.
(\ref{eq:H-transformed}) with respect to $\v v$: $\delta H_0 / \delta
\v v = \delta H_3 / \delta \v v = \rho (\v v + \kappa \v n)$. Such
definition of the current density, albeit odd at first sight, was
anticipated on symmetry grounds by Ambegaokar et al.,
who constructed the Ginzburg-Landau functional for the $A$-phase of
superfluid $^3$He~\cite{AGR}. They argued that the definition of the
current density in general should consist of three types of terms,
the first two of which would reduce to $\rho \v v$ and $\kappa \rho
\v n$ in the case of isotropic superfluid density $\rho_{ij} = \rho
\delta_{ij}$. The third term, proportional to the magnetization
current $\bm \nabla \times \v n$, does not exist in our theory but
might appear when a more general Hamiltonian is employed. It was
mentioned~\cite{AGR} that $\v n$ (axial vector) can appear in the
definition of $\v J$ (polar vector) only if the medium breaks the
inversion symmetry. The Rashba term, being odd under $\bm \nabla
\rightarrow -\bm \nabla$, breaks such symmetry. Additionally, the newly
defined $\v J$ satisfies the requirement of being gauge-invariant,
if we view the Rashba coupling as a non-Abelian gauge field.
\\

\textit{(2) Euler equation}:
\ba
D_t \v v = \v v \times \v b + \v e - \bm \nabla p , \label{eq:EulerEq}
\ea
where
\ba p &=&  {1\over
8} \sum_i (\partial_i \v n)^2 \!+\! \kappa \v n \cdot \v v \!+\!
{\kappa\over 2}\v n \cdot (\bm \nabla \!\times \!\v n) \nn
&& - \frac{\bm \nabla^2 \sqrt{\rho}}{2\sqrt{\rho}} + g_c \rho - g_s  \rho n^2_z .  \nonumber
\ea
The material derivative $D_t = \partial_t + \v v \cdot \bm \nabla$
appears in the above. Compared with the case
in the absence of Rashba coupling~\cite{lamacraft,refael},
only the explicit expression of the quantum pressure $p$ is modified by
$\kappa$. The internal electric ($\v e$) and magnetic ($\v b$)
fields arise from spatial and temporal fluctuations of the
magnetization in the usual way~\cite{refael}: $e_i = -{1\over 2} \v
n \cdot (\partial_i \v n \times
\partial_t \v n)$, $b_i = -{1\over 2}\varepsilon_{ijk} \v n \cdot (\partial_j \v n
\times \partial_k \v n)$.
\\

\textit{(3) Landau-Lifshitz equation}:
\ba\label{eq:spinDynamics}
\rho D_t \v n \! &=& \! \v n \! \times \! \left( {1\over 2}
\partial_{i}(\rho\partial_{i}\v n) \!-\! 2 \kappa \rho \v v
\!-\! \kappa \bm \nabla \times (\rho\v n) \!+\! 2 g_s \rho^2 \v n_z
\right),\nn \ea where $\v n_z = (0,0,n_z)$. Compared to the earlier
expression~\cite{refael}, two new terms appear on the r.h.s. due
to the nonzero $\kappa$ as in Euler equation.

The three equations (\ref{eq:massConservation}), (\ref{eq:EulerEq}),
and (\ref{eq:spinDynamics}) together describe the dynamics of
Rashba-BEC system in hydrodynamic variables.  For their derivations
one can substitute Eq.~(\ref{eq:CP1}) into the time-dependent GP
equation
\ba
i\partial_t \Psi \!=\! \Bigl[ \frac{1}{2}(-i\bm\nabla \!+\!
\kappa\bm\sigma)^2 \!+\! g_c (\Psi^\dag \Psi)\! - \! g_s (\Psi^\dag
\sigma_z \Psi) \sigma_z \Bigr] \Psi \nonumber
\ea
and follow the projection strategies outlined in Ref.~\cite{refael}.
Alternatively, one can start from the hydrodynamic Lagrangian
\ba
{\cal L} = \rho[ -\partial_t \xi  + i \v z^\dag \partial_t \v z
] - H [\rho, \v v, \v n] + \lambda (\v z^\dag \v z - 1), \nonumber
\ea
the constraint $\v z^\dag \v z =1$ being implemented by the Lagrange
multiplier field $\lambda$. The variation of the Lagrangian
$\cal L$ with respect to $\xi$, $\rho$, and $\v z^\dag$, respectively, yields
the mass continuity, Euler equation, and spin dynamics as shown above.
The Lagrange multiplier $\lambda$ term is eliminated in the final equation through
the projection with $\v z_{\mathrm{TR}}$, where $\v z_{\mathrm{TR}} = (\cos(\theta_{\mathrm{TR}}/2)e^{-i\phi_{\mathrm{TR}}/2},
\sin(\theta_{\mathrm{TR}}/2)e^{i\phi_{\mathrm{TR}}/2})^T$ with $\theta_{\mathrm{TR}} = \pi-\theta$ and $\phi_{\mathrm{TR}} = \phi-\pi$,
is the time reversed version of $\v z$ satisfying $\v z^\dag_{\mathrm{TR}} \v z = 0$.

As discussed earlier, the mass conservation law shown in
Eq.~(\ref{eq:massConservation}) suggests the definition of the mass
current density $\v J = \rho (\v v +\kappa \v n)$. In a closed
system, the current density $\v J$ is expected to obey its own
continuity equation $\partial_t J_i + \partial_j \Pi_{ij} = 0$ with
an appropriate momentum flux tensor $\Pi_{ij}$. After some lengthy
algebra, we arrive at such expression as:
\ba
\Pi_{ij} &=& P_{ij} + \kappa Q_{ij}, \\
P_{ij} &=&
-\frac{1}{4}\left( \bm\nabla^2\rho \right)\delta_{ij} + \rho v_i v_j
+ {\rho \over 4}
\partial_i \v n \cdot \partial_j \v n \nn
       &&   +(\partial_i \sqrt{\rho})(\partial_j \sqrt{\rho}), \nn
Q_{ij} &=& 2\kappa\rho\delta_{ij} + \rho(v_i n_j + n_i v_j) \nn
       && - \frac{\rho}{2} \Bigl(
       [\v n \times \partial_i \v n]_j + [\v n \times \partial_j \v n]_i \Bigr) . \nonumber
\ea
We have separated the $\kappa$-dependent part as $Q_{ij}$ in the
above, while the $\kappa$-independent part is denoted as $P_{ij}$. The
last term in $Q_{ij}$ is the well-known spin current in spiral
magnetism~\cite{KNB}. It ought to be emphasized that the continuity
equation applies for $\v J = \rho (\v v + \kappa \v n)$, but not for
$\rho \v v$ alone. Such observation strengthens our earlier claim
about the proper definition of mass current in the Rashba-BEC medium.

Hydrodynamic equations such as derived above can be applied to study
small fluctuations of the known ground states. The three coupled
equations (\ref{eq:massConservation}), (\ref{eq:EulerEq}) and
(\ref{eq:spinDynamics}) have to be solved simultaneously, while also
obeying the constraints implied by the Mermin-Ho
relation~\cite{mermin-ho},
\ba
(\bm\nabla \times \v v)_i &=& \frac{1}{4}\epsilon_{ijk}\v n
\cdot (\partial_j \v n \times \partial_k \v n).
\ea
Most small fluctuation theories such as carried out in
Ref.~\cite{demler} deal with the time-dependent GP equation in the Bogoliubov approach
without explicitly introducing the Mermin-Ho constraint. Here we establish
that the hydrodynamic analysis, carried out in a way to consistently
take care of the Mermin-Ho constraint, yield identical results as the
standard Bogoliubov approach. The great advantage of hydrodynamic analysis,
on the other hand, is that it is much easier to identify the nature
of the given excitation mode as either density, spin, or superfluid
velocity oscillations, or some mixture thereof. Below we examine two
known ground states of the Rashba-BEC Hamiltonian $H= H_0 +
H_\mathrm{I}$. The first one, called the plane-wave state, is
obtained for $g_s <0$ and reads $\Psi_0 = (1/\sqrt{2}) e^{i\kappa x
}(1, -1 )^T$~\cite{zhai}. The other stripe state exists for $g_s
>0$, and reads $\Psi_0 = ( \cos \kappa x, -i\sin \kappa
x)^T$~\cite{zhai}. The two ground states become degenerate at the
critical point $g_s=0$, with the energy $E=-\kappa^2/2+g_c$. In
typical experimental situations we always have $g_c \gg |g_s |$,
hence we take $g_s=0$ in the following to simplify the analysis.

\begin{figure}[tbph]
\centering
\includegraphics[width=85mm]{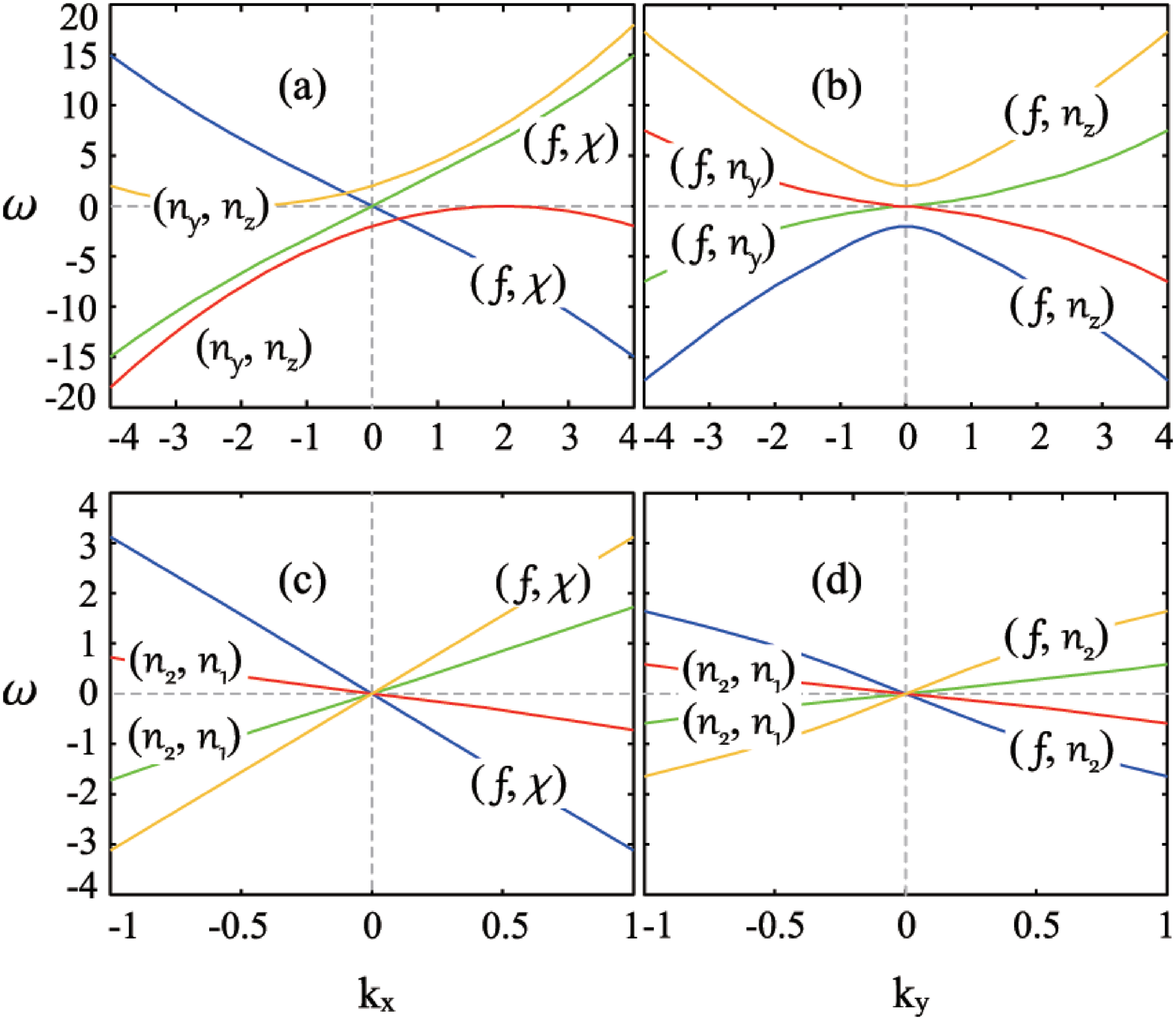}
\caption{(color online) Excitation modes of plane-wave (upper
panels) and stripe (lower panels) state along the $k_x$ (left column,
$k_y = 0$) and $k_y$ (right column, $k_x=0$) directions. Rashba
strength $\kappa = 1$ sets the unit for the $x$-axis and $g_c =10$.
Only the four gapless branches are shown in the lower panels for the
stripe state. We choose $g_s = 0$ for all cases. Shown inside the
parentheses are the two components contributing the most and the
second most to each eigenvector. }\label{fig:eM}
\end{figure}

The plane-wave state is characterized by $\rho_0=1$, $\v
v_0=(\kappa,0,0)$, and $\v n_0 =(-1,0,0)$. It is a state with
ferromagnetic spin order $\v n_0$ and nonzero superflow $\v v_0$,
which however combine to produce zero mass current $\v J_0 = \v v_0 +
\kappa \v n_0 = 0$. Small fluctuations around each variable is
parameterized as $\rho=\rho_0+\chi$, $\v v= \v v_0 + \bm \nu$, $\v
n=\v n_0 + \eta_y \v e_y + \eta_z \v e_z$, and inserted into the
hydrodynamic equations (\ref{eq:massConservation}),
(\ref{eq:EulerEq}) and (\ref{eq:spinDynamics}) to first order in
$\chi, \bm \nu, \eta_y, \eta_z$. Assuming the two-dimensional
fluid, fluctuation in the $z$-direction will be ignored. The
effective electric and magnetic fields do not contribute to the Euler
equation in the linear analysis. Furthermore, Mermin-Ho relation as
applied to the plane-wave state yields the constraint $\bm \nabla
\times \bm \nu = 0$, which is solved by writing $\bm \nu = \bm \nabla
f$. In $\v k$-space, the scalar functions $\chi$ and $f$, and two
components of the spin fluctuation $\eta_y$ and $\eta_z$ obey a
four-dimensional matrix equation of motion, $ \mathcal{M} (\chi,
f, \eta_y, \eta_z)^T = 0$, where $\mathcal{M}$ equals
\ba
\bpm -\omega & i(k^2_x\!+\!k^2_y) & \kappa k_y & 0 \\
-i\frac{\v k^2}{4}\!-\!i g_c & -\omega & 0 & -\frac{\kappa}{2} k_y \\
\kappa k_y & 0 & 2\kappa k_x\!-\!\omega & -i\frac{\v k^2}{2} \!-\!
2\kappa^2 i \\ 0 & -2\kappa k_y & i\frac{\v
k^2}{2}\!+\!2\kappa^2 i & 2\kappa k_x \!-\!\omega \epm. \nonumber\ea
Diagonalizing the matrix, one obtains the
dependence of excitation mode $\omega$ on $\v k$ as shown in
Figs.~\ref{fig:eM}(a) and \ref{fig:eM}(b). For the parameters we
choose, we can observe the branches of one shifted mode in $k_x$
direction and one gapped mode in $k_y$ direction. These features
coincide with the ones obtained in Ref.~\cite{biao-wu}, which was
based on the analysis of time-dependent GP equation in the
Bogoliubov approach. Furthermore, we label in the dispersions the two components
which contribute the most and the second most to each eigenvector. For instance,
the shifted modes in Fig.~\ref{fig:eM}(a) come from the spin fluctuation, while
the linear sound modes are due to fluctuations of the total density and the overall phase.
Similar analysis applies to Fig.~\ref{fig:eM}(b).
Hydrodynamic theory indeed enables us to directly identify different modes more easily
than Bogoliubov approach.
\\

The stripe solution yields $\rho_0=1$, $\v v_0 =0$, and $\v n_0 =
(0, -\sin \theta, \cos \theta)$, where $\theta = 2\kappa x$. The
spin order is a spiral with the plane of spin rotation orthogonal to
the propagation vector $\v q = 2\kappa \v e_x$. Only the
fluctuations of $\v n$ need to be parameterized differently from the
plane-wave case, as $\v n = \v n_0 + \eta_1 \v e_1 + \eta_2 \v e_2$,
where $\v e_1 = (1,0,0)$, $\v e_2 = (0, \cos \theta, \sin\theta)$, and $\v n_0$
form the three orthogonal axes in the rotating Frenet-Serret
frame~\cite{lamacraft}. Contrary to the
plane-wave case, we have a nonzero effective electric field
contribution to the Euler equation as $e_x = -\kappa \dot{\eta}_1$ and
correspondingly, the Mermin-Ho constraint in the case of stripe
state is solved by $\nu_x + \kappa \eta_1 = \partial_x f, \nu_y
=\partial_y f$. Inserting them back into the hydrodynamic equations
yields
\ba \label{eq:stripeP}
-\omega \chi + \frac{i}{2}\kappa k_y \chi^- + i\v k^2 f +
\frac{\kappa}{2} k_y \eta^+_2 &=& 0, \nn
(\frac{i}{4}\v k^2-i g_c)\chi - \omega f + \frac{i}{2}\kappa k_y
f^- -\frac{\kappa}{4}k_y \eta^+_1 &=& 0, \nn
-\kappa k_y f^+ -\omega\eta_1 - \frac{i}{2}\kappa k_y \eta^-_1 +
\frac{i}{2}\v k^2 \eta_2 &=& 0, \nn
\frac{\kappa}{2} k_y \chi^+ - \frac{i}{2}\v k^2 \eta_1 -2i
\kappa^2 \eta_1 -\omega \eta_2 + \kappa k_x \eta_2
  -\frac{i}{2} \kappa k_y \eta^-_2 &=& 0, \nonumber
\ea
where $m^{\pm} = m(\v k - \v q) \pm m(\v k+ \v q)$
($m=\chi, f, \eta_1, \eta_2$) and un-superscripted variables are
at momentum $\v k$.
Coupling of the $\v k$-modes displaced by $\pm \v q$ is a consequence
of the ground state being a modulated state, thus no way to write the excitation
equations into the $4\times 4$ matrix form as in the plane wave case. Numerical
diagonalization yields the spectrum shown in Fig.~\ref{fig:eM}(c) and
\ref{fig:eM}(d). Analogous ``helimagnon band" was identified in a
spiral magnet MnSi~\cite{rosch}. For the parameters we choose, four linear
sound modes with mode contributions as labeled in Fig.~\ref{fig:eM}(c) and
\ref{fig:eM}(d) exist near $k_x = k_y =0$, which
is different from the plane-wave case. Bogoliubov analysis of the GP
equation yields identical spectra for the stripe state as well.

In summary, we have derived the mathematical mapping of the Rashba
spin-orbit-coupled BEC Hamiltonian in terms of hydrodynamic
variables. The mapping highlights the connection of the present
system to the chiral magnets. The hydrodynamic equation for mass conservation,
as well as Euler and Landau-Lifshitz
equations for superfluid velocity and spin dynamics, respectively,
are derived. The mass current contains an extra magnetization term due to the Rashba coupling.
As applications of hydrodynamic approach, we studied
the elementary excitation modes of the plane-wave and stripe states
of Rashba-BEC, featuring the different behaviors of fluctuations around
the two degenerate ground states and the coupled behavior of spin
and superflow velocity in the presence of Rashba coupling.

H. J. H. is supported by NRF grant (No. 2011-0015631). Useful
discussion with Hui Zhai at the early stage of research is
gratefully acknowledged.

\end{document}